\documentclass[aps,12pt,final,notitlepage,oneside,onecolumn,nobibnotes,nofootinbib,superscriptaddress,noshowpacs,centertags]{revtex4}
\usepackage{amsmath,amssymb,epsfig,placeins,subfigure,wrapfig,indentfirst,makeidx}
\usepackage[T2A]{fontenc}
\usepackage[utf8x]{inputenc}
\usepackage{ucs}
\usepackage{amsmath} 
\usepackage{mathtext}
\usepackage{amsfonts}
\usepackage{upgreek}
\usepackage{amsfonts,amssymb}
\usepackage{color}
\usepackage{graphicx}
\usepackage{mathrsfs}
\usepackage{dsfont}

\begin{document}

\title{Neural Network Astronomy as a New Tool for Observing Bright and Compact Objects}

\author{Alexander Shatskiy}
\affiliation{shatskiyalex@gmail.com}
\author{Ivan Evgeniev}

\begin{abstract} 
We propose a new method for solving an important problem of astronomy that arises in observations with ultrahigh-angular-resolution interferometers. 
This method is based on the application of the theory of artificial neural networks. 
We propose and compute a multiparameter model for a celestial object like Sgr A*. 
For this model we have numerically constructed a number of probable images for neural network training. 
After neural network training on these images, the quality of its operation has been tested on another series of images from the same model. 
We have proven that a neural network can recognize and classify celestial objects (also obtained from interferometers) virtually no worse than can be done by a human.
\end{abstract}

\maketitle

\section{INTRODUCTION}
\label{introduction}

Interferometers with an angular resolution of less than 50 microarcseconds ($\mu$as) are required to observe the characteristics of objects whose angular sizes are comparable to the angular sizes of the horizon diameter for supermassive black holes. 

At the same time, the observations of many objects in the Universe with single telescopes, even space ones, do not allow their structure to be investigated due to their angular sizes being small. 
The angular resolution of modern long-baseline interferometers in the optical, infrared, and radio bands (the latter are
the so-called very-long-baseline interferometers, VLBI) approaches 10 $\mu$as
\footnote{${1\mu as \approx 4.8\cdot 10^{-12} rad}$.} – see~\cite{VLBIbook, KM, Q_rev, Q_space, Lu, RA}.

We are talking primarily about the observations of Sgr A* (the shadow angular diameter \mbox{$\approx$ 60 $\mu$as}) and the nucleus of M87 (the shadow angular diameter \mbox{$\approx$ 40 $\mu$as}).

As is well known (see, e.g.,~\cite{VLBIbook}), interferometers do not see the images of objects, they see (measure) the
amplitude ${A\mbox{(u,v)}}$ and phase ${\Phi\mbox{(u,v)}}$ of the complex visibility function 
${V\mbox{(u,v)} := A\cdot\exp(i\Phi)}$ of these objects:

\begin{eqnarray}
V\mbox{(u,v)} = \int\limits\int\limits I\mbox{(x,y)} \exp\left[ -2\pi i \mbox{(xu+yv)}/\lambda \right] \, d\mbox{x}\, d\mbox{y} \, , \label{intr_V_uv}\\
I\mbox{(x,y)} = \int\limits\int\limits V\mbox{(u,v)} \exp\left[ 2\pi i \mbox{(xu+yv)}/\lambda \right] \, d\mbox{u}\, d\mbox{v} \, .\label{intr_I_xy}
\end{eqnarray} 
Here, ${I\mbox{(x,y)}}$ is the intensity function of the observed object in angular coordinates and $\lambda$ is the wavelength
at which the observation is carried out. 
The coordinates $\mbox{(u,v)}$ are actually the two-dimensional coordinates of the interferometer baseline and are called the coordinates on the $\mbox{(u,v)}$ plane.

Observations on very small angular scales can be performed only with very-long-baseline interferometers: more than 
${5\cdot 10^9\lambda}$, see~\cite{VLBIbook}. 
The main problem here is to measure the phase at a sufficiently large path difference to the interferometer telescopes. 
In this case, the minimum required accuracy is a quarter of the wavelength, ${\lambda/4}$, which corresponds to the phase measured with an accuracy much less than ${\pi/2}$. 
For infrared and optical interferometers this requirement is, in principle, satisfiable, but the spectral resolution
of the telescopes (${\lambda /\Delta\lambda}$) in such interferometers is of the order of or less than unity. 
This leads to strong blurring of the interference pattern. 
Therefore, without precise knowledge of the spectra of image details, the signal processing is meaningless in the received
band ${\Delta\lambda}$. 
Unfortunately, we can measure only the total spectrum of the entire image, but we cannot measure the spectra of image details. Therefore, we can quantitatively process the interferometric data only in those bands where the monochromatic approximation for
the spectral resolution is admissible: ${\lambda /\Delta\lambda >>1}$.

At present, this corresponds to telescopes operating at wavelength longer than $\sim$1 mm.

\section{PECULIARITIES OF INTERFERENCE ASTRONOMY}
\label{interastron}

Unfortunately, most of the interferometers with an ultrahigh angular resolution (i.e., with very long baselines) cannot measure the phase of the visibility function, they record only its amplitude. 
The terms “Correlated Flux Density” (CFD) are commonly used instead of the word “amplitude” in Russian and English language literature, respectively; below we will everywhere use the abbreviation CFD. 
The phase of the visibility function is irretrievably lost when the correlation length is exceeded and due to atmospheric (and other) noise.
Nevertheless, even the knowledge of the CFD alone can clarify much about the image of an object and its properties.

% --------------------------------------------- fig 1
\begin{figure}
[h]
\centering
\begin{minipage}[h]{0.90\linewidth}
\includegraphics[width=0.90\textwidth]{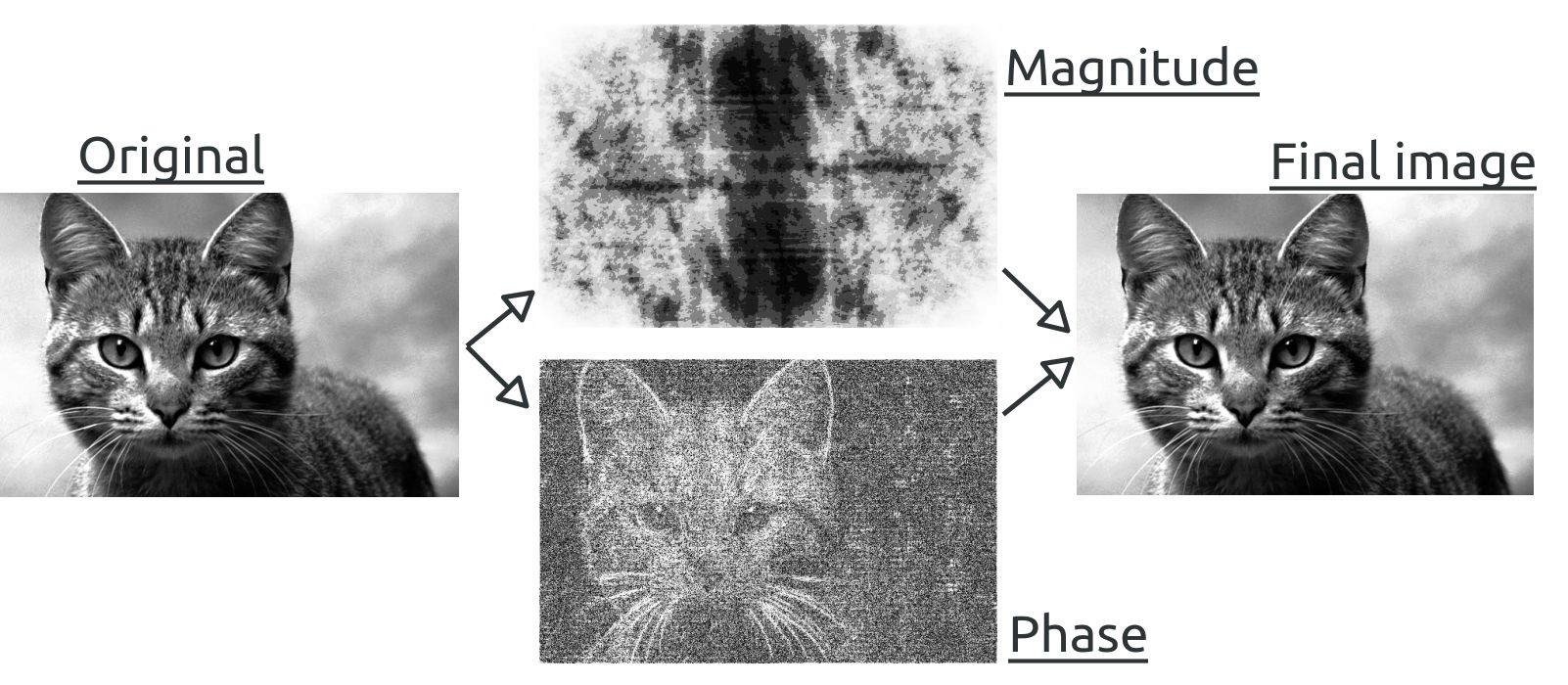}
\caption{Direct and inverse Fourier transforms for the image of a cat. 
The image for the amplitude (magnitude) was obtained as the reconstructed (by the inverse Fourier transform) image with a phase equal to unity and the cat’s amplitude, while the image for the phase was obtained as the reconstructed image with an amplitude equal to unity and the cat’s phase.
}
\label{R1}
\end{minipage}
% ---------------------------------------------

% --------------------------------------------- fig 2
\vfill
\begin{minipage}[h]{0.90\linewidth}
\includegraphics[width=0.90\textwidth]{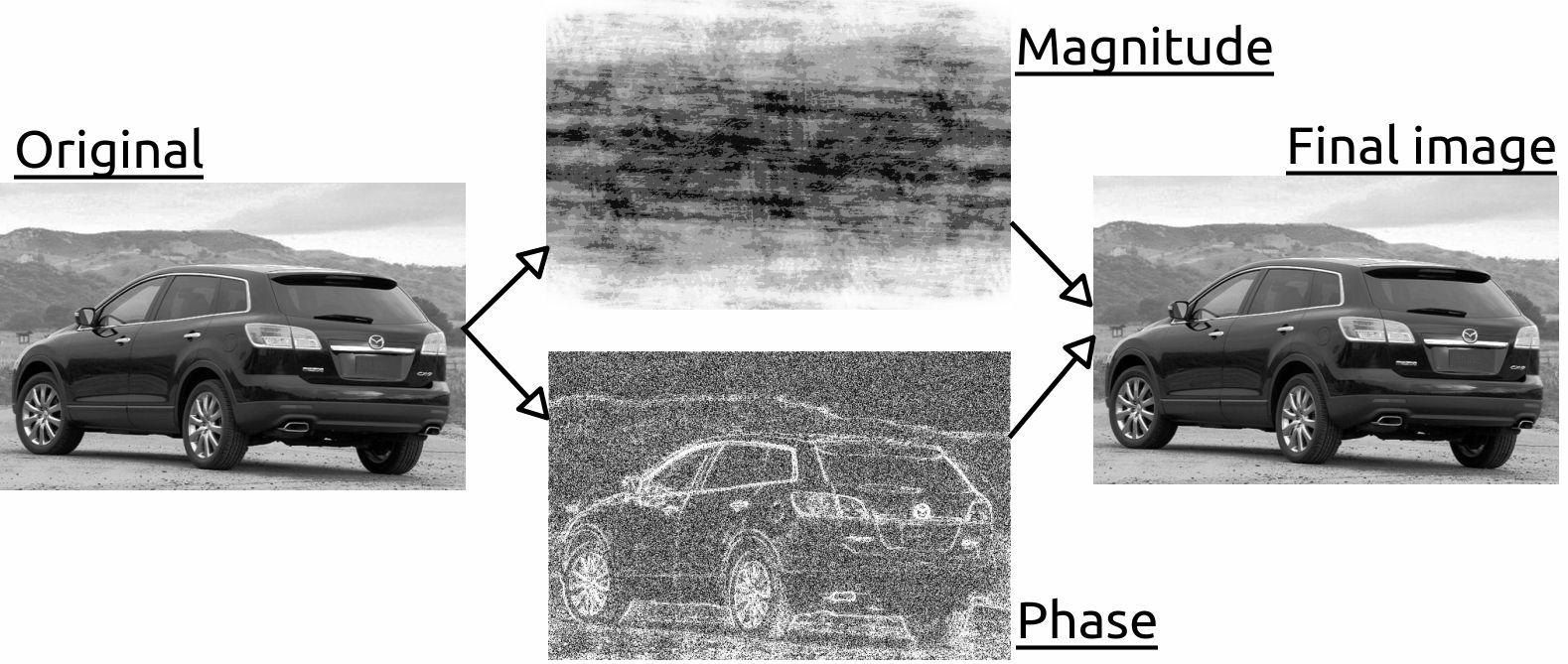}
\caption{Same as Fig.~\ref{R1} for the image of a car.}
\label{R2}
\end{minipage}
\end{figure}
% -----------------------------------------------

% --------------------------------------------- fig 3
\begin{figure}
\centering
\includegraphics[width=0.90\textwidth]{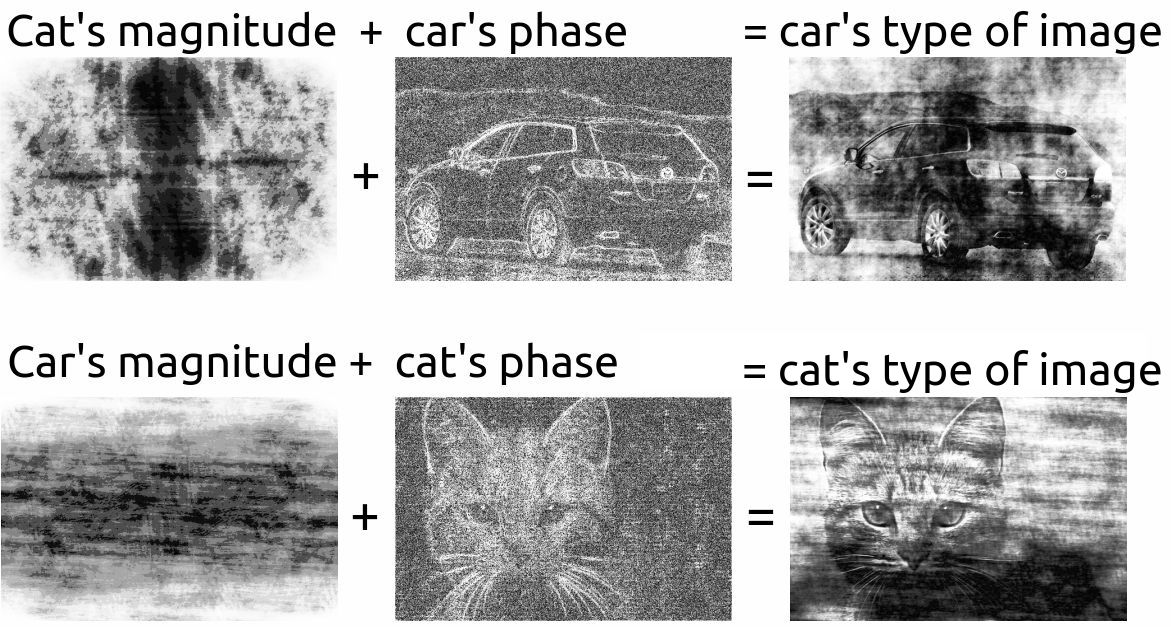}
\caption{(a) Cat’s magnitude + car’s phase; the reconstructed image thus resembles the car. 
(b) Car’s magnitude + cat’s phase; the reconstructed image thus resembles the cat.}
\label{R3}
\end{figure}
% -----------------------------------------------

What can be said about the image without knowing the phase? 

Quite a few papers have been written on this subject (see, e.g.,~\cite{Coles_Chiang, Chiang}).

Figures (\ref{R1}-\ref{R3}) demonstrate the direct and inverse Fourier transforms for two images (in shades of gray):
a cat and a car. 
As is clear from the figures themselves and the captions to them, the basic information that allows the human to identify the image is contained in the phase $\Phi$. 
However, this, per se, does not imply that less information is contained in the CFD than in the phase $\Phi$. 
For some reasons, the human brain and vision are unable to distinguish the information contained in the CFD. 
The human distinguishes only the information contained in the phase $\Phi$.

Knowing only the CFD, until recently we have had only one opportunity to determine the image characteristics — to compare the distribution of the observed CFD with its model values in the form of some two-dimensional function dependent on a specific set of parameters. 
Thus, we could choose the best model image corresponding to a specific set of these parameters (the best vector in the parameter space). 
There are many mathematical criteria for the degree of similarity (and dissimilarity) between functions. One of them is the well-known chi-square ${(\xi^2)}$ test:

\begin{eqnarray}
\xi^2 = \sum\limits_{i} \frac{(A_{model,i} - A_{obs,i})^2}{\Delta A_{obs,i}^2}
\label{intr_xi2}
\end{eqnarray} 
Here, the subscripts $\mbox{«model»}$ and $\mbox{«obs»}$ correspond to the model and observed amplitudes, respectively; 
the summation is over all of the image points the data on which is available to the observer.

The chi-square test is fairly simple and reliable, but it requires an exponentially long computation time at a large number of model parameters.

Suppose that we have an $N$-parameter model of the CFD as a two-dimensional function on the $\mbox{(u,v)}$ plane.

Let there also be $T$ observation points on the $\mbox{(u,v)}$ plane at which we will compare the observation and model results. 
In our $N$-dimensional parameter space we must choose a cube and divide each edge of this
cube into $K$ parts (in principle, this number can be different for each edge). 
The total number of points (cells) at which we must calculate the chi-square is then ${K^N}$ and to calculate the chi-square, we must multiply ${K^N}$ by ${5T}$ to obtain the total number of necessary arithmetic operations. Actually, however, we will need even more computational resources, because each model point is usually an arithmetic expression composed of special functions (usually Bessel functions) and their calculation also takes time. 
For example, the total number of arithmetic operations $X$ for N${N=5}$, ${K=100}$ и ${T=100}$
 is 
${X=5\cdot 10^{12}}$ (not counting the computations of the model points themselves). 
We chose this example not accidentally: it corresponds to the finished computation of models by astrophysicists from the group of the Event Horizon Telescope~\cite{crescent}.

The question arises as to why the standard methods of finding the global minimum of an $N$-dimensional
function in an $N$-dimensional parameter space cannot be used.

Unfortunately, the direct and obvious methods of optimizing such calculations, for example, the gradient descent method in the functional (\ref{intr_xi2}), are inapplicable in this case. 
This is because calculations of this kind by the standard methods lead to one of the local minima, while we need the global minimum (or the local minimum, but near the global one). 
Since the local minima become exponentially many at large $N$, there is very little chance to find the global one among them. 
It is also worth noting that, despite the fact that the gradient descent method is inapplicable to finding the minimum of the functional (\ref{intr_xi2}), this method is commonly used in neural network training—the error backpropagation algorithm (which will be discussed below) allows the gradient of the loss function in weights to be calculated efficiently.

\begin{figure}
\centering
\includegraphics[width=0.40\textwidth]{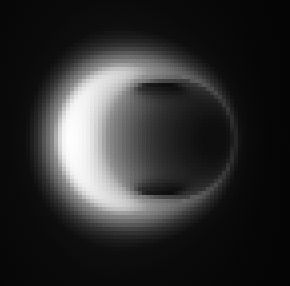}
\caption{The blurred (Gaussian) crescent model that well approximates the shadow from a black hole.}
\label{R4}
\end{figure}

The CFD properties for objects that could be black holes were investigated in~\cite{Shatskiy2015}. 
The shadows that the gravity of a black hole leaves against the radiation background whose source is the matter surrounding the black hole are common to the black hole models. 
The source of this radiation can be the matter constituting an accretion disk around the black hole or this radiation is simply the background one, including the cosmic microwave background. 
In any case, the shape of this shadow in the pattern of the radiation recorded by an observer is very characteristic. 
An excellent approximation for this shape is the crescent model or the Gaussian blurred crescent model (Fig.~\ref{R4}) (see, e.g.,~\cite{crescent}).

The crescent model is very simple and efficient for numerical simulations. 
It has only five numerical parameters and, as has been said above, the number of 

arithmetic operations for this model is ${\sim 10^{13}}$. 
For modern computers this is a feasible task.

But what to do if our model is more complex and has, for example, 11 parameters? 
In this case (at ${K=100}$ and ${T=100}$), the number of arithmetic operations reaches ${\sim 10^{25}}$. 
This is already beyond the limits of even modern supercomputers, while the number of parameters $N$ in more realistic models can be much greater (about a hundred).

What to do in this case?

\section{NEURAL NETWORK ASTRONOMY}
\label{neuralastron}

The science of neural network astronomy is completely new and this definition was introduced by one of the authors of this paper. Let us first clarify what artificial neural networks (or neural nets for short) to be talked about below are.

An artificial neural network is a mathematical model (and its software or hardware implementation) constructed on the principle of organization and operation of biological neural networks — the networks of nerve cells in a living organism. 
This concept arose in studying the processes occurring in the brain and in attempting to model these processes.

Forecasting, approximation, clustering, data compression, object localization, and partitioning a specified set of objects into classes (and occasionally also into subclasses) are practically solvable problems for many (artificial) neural nets. 
We will be interested precisely in partitioning a specified set of objects into classes, i.e., briefly speaking, the set classification. 
For this purpose, we train the neural net—randomly chosen objects from the training set are fed to its input in turn. 
At the output the neural net gives a classifier vector. 
Ideally, for a completely trained neural net this vector is filled with 0s and one 1—the position number of this 1 corresponds to the class number of the current object at the input. 
In reality, however, this vector consists of noninteger numbers (float) and is normalized to unity. 
The error with which the neural net determined the sought-for class can be derived from this vector. 
With the error backpropagation algorithm this error propagates from the end of the network to its beginning and a correction of the neural net weighting factors (which form its basis) occurs in the process.
This process is repeated until the error at the output becomes sufficiently small and the neural net then is then ready for operation.

Artificial neural networks can be read about in more detail in~\cite{kruglov_borisov, LeCun, nn1, cnn1}. 
The neural nets used in this paper can be read about in the series of 
papers~\cite{Darknet1, Darknet2, Darknet3, Darknet4, Darknet5, Darknet6, Darknet7} and~\cite{Darknet0}.

As has been mentioned above, for some reasons, the human brain and vision are unable to distinguish the information contained in the CFD, although quite a lot of information about the object is also contained there. 
The idea of this paper is to train the neural net on the objects that are the amplitudes of the image Fourier transforms, i.e., the CFD. 
As will be seen below, this turns out to be a completely solvable problem.

% --------------------------------------------- fig 4
% -----------------------------------------------

Thus, we can train the neural net on various classes in the chosen model — for a wide range of extended astronomical objects or a group of point objects.

The neural net can be trained equally well no matter how complex the models of these objects are. 
This unique possibility solves the problem formulated in the previous section.

\section{A MODEL FOR THE NEURAL NET}
\label{model}

For this paper we chose a relatively simple model containing 11 parameters. 
Nevertheless, as has been mentioned above, this model cannot be computed by the standard methods even with supercomputers. 
We will show that neural nets are capable of classifying such a model with a probability close to ${100\%}$.

As the basis for the model we will use the crescent model — it includes five parameters: 
the radius of the outer circle, the radius of the inner circle, two coordinates of the center of the inner circle (relative to the center of the outer circle), and the crescent brightness.

\begin{figure}
\centering
\includegraphics[width=0.90\textwidth]{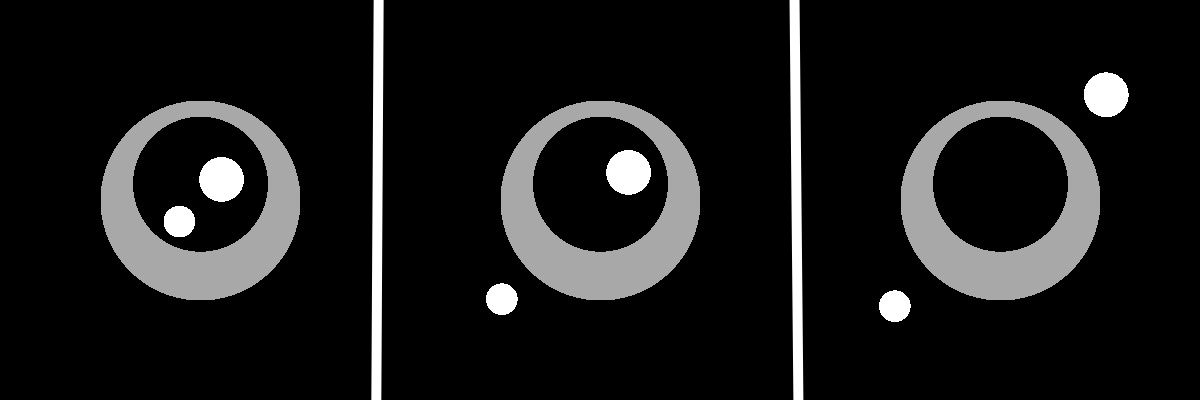}
\caption{The 11-parameter model for neural net training on three classes: 
(a) the zeroth class (no stars outside the crescent’s circle), 
(b) the first class (one star outside the crescent’s circle), 
and (c) the second class (two stars outside the crescent’s circle).}
\label{R5}
\end{figure}

Furthermore, we add two more point objects\footnote{In the model itself we add not bright points, but bright circles
whose radius is proportional to the star’s model brightness to the pictures, because the numerical simulations of point objects
yield inadequate results compared to the simulations of extended objects. 
Furthermore, in real telescopes stars give images precisely in the form of circles (rather than points); 
the brighter the star, the bigger the circle. 
In the subsequent Fourier transforms we simulate the same stars precisely as points (delta functions).} (two stars) to the model, each of which has three parameters: 
their coordinates and brightness~\ref{R5}.

According to the study in~\cite{Shatskiy2015}, the stars inside the crescent’s outer circle correspond to the model of a black-white hole or a wormhole, while the stars outside the crescent’s outer circle correspond to stars against the background of a black hole (or the stars gravitationally lensed by this black hole).

% --------------------------------------------- fig 5
% -----------------------------------------------

The training can be done both on the original model images (which, in reality, we do not have, as was shown in Section~\ref{interastron}) and on the CFD images corresponding to the original ones.

Furthermore, we can also attempt to perform the inverse Fourier transforms for the CFD images by taking a constant 
two-dimensional function instead of the missing phase and, as a result, to obtain another set of images (also partitioned into classes) and thereafter to train the neural net on this new set of images.

\section{NEURAL NET TRAINING}
\label{lerning}

We chose the YOLOv3 neural net from the Darknet package~\cite{Darknet0} (see~\ref{app1}) for training. 
This package belongs to the open software with the BSD license. 
Such a choice is dictated by the fact that, first, this neural net can distinguish the classes on which it was trained with a fairly high quality. 
Second, this neural net can also construct the frames around the objects found on the image—this may be required to identify the subclasses (subsets) on the sets of objects (in subsequent studies). 
Third, this neural net also gives the probability of the object determined by it (outlined by a frame, with the assignment of an identifier to it). 
Thus, we can cut off those objects whose detection probability is not high enough for us.

We also attempted to train much simpler (shallow) neural nets on the same set of images. 
These shallow neural nets were only three convolutional layers and one fully connected (perceptron) layer at the output.
However, this configuration did not allow a class recognition probability higher than 70\% to be obtained.

We decided to test the neural net training quality for three different sets of pictures: 
1) the set of original model pictures, 2) the set of CFD pictures obtained by an analytical Fourier transform from the original
ones (by taking into account the use of a delta function in the Fourier transform for the models of stars), and 3) the set of pictures obtained from the CFD pictures by a numerical inverse Fourier transform using a constant phase function (hereafter the CFD2DFT pictures for short).

To analytically obtain the CFD pictures from specified 11 parameters, we used the formula for the CFD from~\cite{Shatskiy2015}:

\begin{eqnarray}
A_{crescent+points}{\rm (u,v)} = \nonumber\\
=
\left|\frac{B_0}{\eta}\left[ r_{out}\, J_1(\eta r_{out}) - e^{-2\pi i 
{\rm (x_c u+y_c v)}/\lambda}\, r_{in}\, J_1(\eta r_{in}) \right] 
+ \sum\limits_{j} B_j e^{-2\pi i (x_j {\rm u}+y_j {\rm v})/\lambda} \right| \quad 
\label{furye}\end{eqnarray}
Here, ${\eta := \sqrt{\rm{u^2+v^2}}}$, $B_0$ is the crescent’s brightness, $r_{in}$ and $r_{out}$ are its inner and outer radii, $x_c$ and $y_c$ are the coordinates of the center of the crescent’s inner circle
(relative to the outer one), $B_j$ is the brightness of the star with index $j$, $x_j$ and $y_j$ are its coordinates, $J_1$ is a
first-order Bessel function, and $\lambda$ is the wavelength at which the observations are carried out.

We prepared all pictures and the transformations on them using the OpenCV package in C++.

For each of these three sets of pictures we prepared a training set of 3000 pictures—1000 pictures for each of the three classes (0, 1, 2). The neural net was trained on each of these sets (3000 pictures each).

In addition, we prepared 600 more pictures for each set (200 pictures in each class) for the tests of the trained neural nets. All these pictures differ randomly from one another in 11 model parameters. 
Furthermore, we made sure that two stars in the pictures were not superimposed on each other (too closely) and did not lie too
close to the boundary of the crescent’s outer circle in order that the classes differ clearly from one another. 

% --------------------------------------------- fig 6
\begin{figure}
[h]
\centering
\begin{minipage}[h]{0.80\linewidth}
\includegraphics[width=0.80\textwidth]{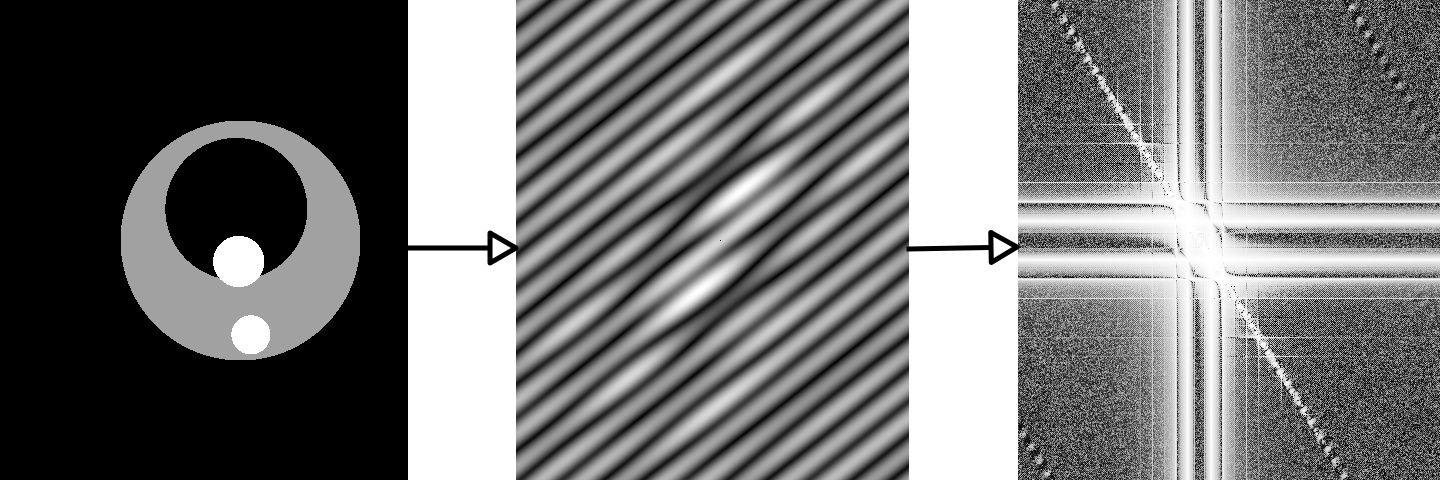}
\end{minipage}
% --------------------------------------------- fig 7
\vfill
\begin{minipage}[h]{0.80\linewidth}
\includegraphics[width=0.80\textwidth]{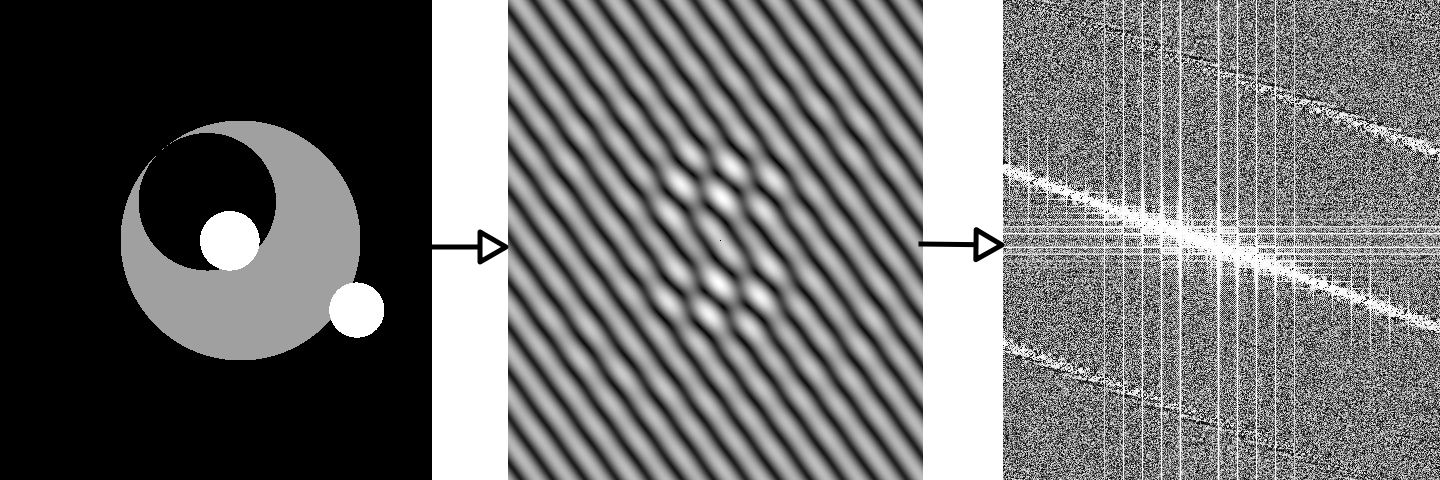}
\end{minipage}
% --------------------------------------------- fig 8
\vfill
\begin{minipage}[h]{0.80\linewidth}
\includegraphics[width=0.80\textwidth]{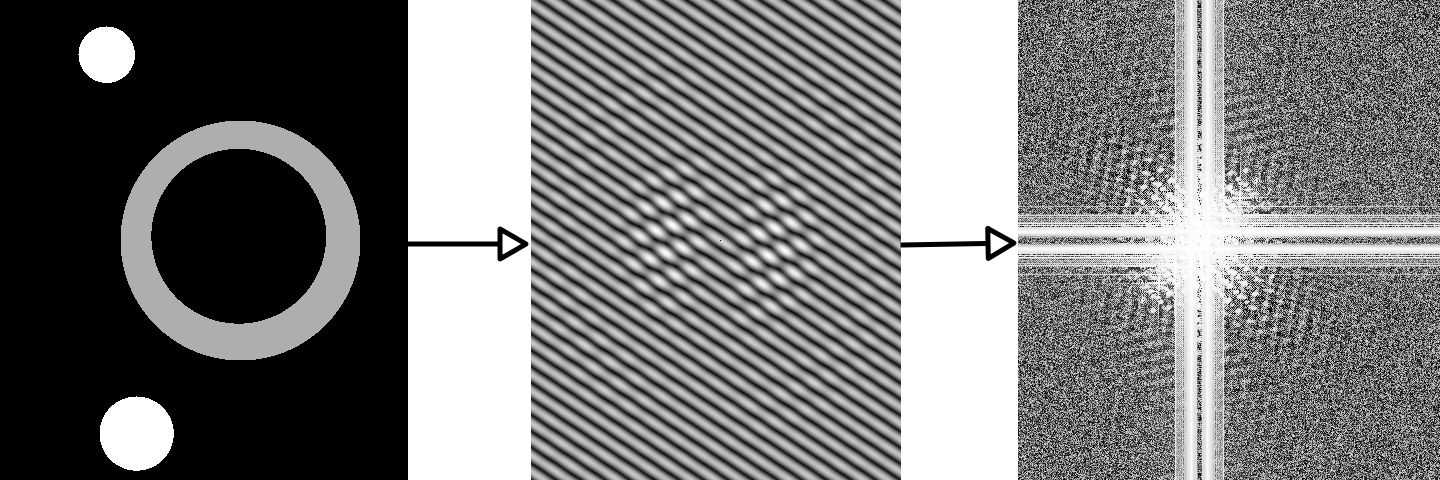}
\end{minipage}
\caption{The original pictures are on the left, the CFD pictures obtained with the same 11 parameters as those for the original ones are at the center, and the pictures obtained from the CFD pictures by the inverse Fourier transform using a constant phase function (CFD2DFT pictures) are on the right. 
The upper, middle, and lower rows are for the zeroth, first, and second classes, respectively.}
\label{R6_8}
\end{figure}
% -----------------------------------------------

Examples of such pictures (for each of the three classes) are presented in~\ref{R6_8}.

\section{NEURAL NET TRAINING RESULTS}
\label{results}

Among other characteristics and parameters (to be discussed below), the neural net gives the probability «${prob}$» of recognizing the object classified by it (see the~\ref{app1}). 
Denote the minimum threshold for this probability by $prob_{min}$.

Below there is yet another {\bf probability of the correct recognition of classes} defined as the fraction of correct answers among all answers of the neural net on a sample of 600 test pictures (200 pictures in each class); 
we will denote this probability by $\varrho$.

At ${prob_{min} = 0.1}$ the probabilities $\varrho$ were found to be the following:
\newline 
– for the original pictures ${\varrho\approx 98.83\%}$,
\newline 
– for the CFD pictures ${\varrho\approx 97.67\%}$,
\newline 
– for the CFD2DFT pictures ${\varrho\approx 92.20\%}$.

In this case, all pictures were recognized (either correctly or not), i.e., $prob$ for all 600 pictures turned out to be greater than $prob_{min}$. 
An analysis of the errors showed that the neural net made a mistake only in the cases where two stars were partially superimposed on each other, i.e., the neural net took them as one star (while it could not find the second star).

If, however, we want to reject unreliable results, then the minimum threshold $prob_{min}$ should be raised.

At ${prob_{min} = 0.9}$ the probabilities were found to be the following:
\newline 
– for the original pictures ${\varrho\approx 99.32\%}$, 15 pictures were unrecognized;
\newline 
– for the CFD pictures ${\varrho\approx 98.97\%}$, 20 pictures were unrecognized;
\newline 
– for the CFD2DFT pictures ${\varrho\approx 95.19\%}$, 122 pictures were unrecognized.

At ${prob_{min} = 0.95}$ the probabilities were found to be the following:
\newline  
– for the original pictures ${\varrho\approx 99.30\%}$, 28 pictures were unrecognized;
\newline  
– for the CFD pictures ${\varrho\approx 99.13\%}$, 28 pictures were unrecognized;
\newline  
– for the CFD2DFT pictures ${\varrho\approx 95.57\%}$, 171 pictures were unrecognized.

\section{ADDING NOISE TO THE PICTURES}
\label{noise}

In real measurements with interferometers an observer does not have a completely filled $\mbox{(u,v)}$ plane, there are only isolated curves on this plane. 
Therefore, to obtain the result, the unfilled part of the $\mbox{(u,v)}$ plane is approximated based on the available observational data (lines on the $\mbox{(u,v)}$ plane). 
In this case, the errors related to this approximation inevitably arise. 
Furthermore, the errors related to the noise superimposed on the observational data are added.

To estimate the influence of all these errors and noise on our results, we artificially “spoiled” the generated CFD pictures. 
We did this by superimposing a blur filter on the pictures that blurred (or smoothed) the picture with a specified parameter of the blurring region. 
In this case, the neural net was not retrained on the blurred pictures.

The sizes of all our pictures are 480x480 pixels.

For a blurring region with sizes of 3x3 pixels:
\newline  
– for ${prob_{min} = 0.1}$ we have ${\varrho\approx 98.17\%}$, all pictures were recognized;
\newline  
– for ${prob_{min} = 0.9}$ we have ${\varrho\approx 98.80\%}$, 15 pictures were unrecognized;
\newline  
– for ${prob_{min} = 0.95}$ we have ${\varrho\approx 99.13\%}$, 27 pictures were unrecognized.

For a blurring region with sizes of 5x5 pixels:
\newline  
– for ${prob_{min} = 0.1}$ we have ${\varrho\approx 96.83\%}$, all pictures were recognized;
\newline  
– for ${prob_{min} = 0.9}$ we have ${\varrho\approx 98.78\%}$, 27 pictures were unrecognized;
\newline  
– for ${prob_{min} = 0.95}$ we have ${\varrho\approx 99.10\%}$, 42 pictures were unrecognized.

For a blurring region with sizes of 10x10 pixels:
\newline  
– for ${prob_{min} = 0.1}$ we have ${\varrho\approx 93.67\%}$, all pictures were recognized;
\newline  
– for ${prob_{min} = 0.9}$ we have ${\varrho\approx 96.55\%}$, 50 pictures were unrecognized;
\newline  
– for ${prob_{min} = 0.95}$ we have ${\varrho\approx 96.82\%}$, 65 pictures were unrecognized.

For a blurring region with sizes of 20x20 pixels:
\newline  
– for ${prob_{min} = 0.1}$ we have ${\varrho\approx 65.61\%}$, 1 picture was unrecognized;
\newline  
– for ${prob_{min} = 0.9}$ we have ${\varrho\approx 71.69\%}$, 155 pictures were unrecognized;
\newline  
– for ${prob_{min} = 0.95}$ we have ${\varrho\approx 74.18\%}$, 205 pictures were unrecognized.

Hence we see that this blurring (or smoothing) of pictures, as would be expected, has a negative effect on the results of the correct recognition of classes by our neural net. 
However, the final results still turn out to be quite acceptable for using neural nets as artificial intelligence for recognizing astronomical objects. 
We hypothesize that such an insignificant influence of noise on the results stems from the fact that, basically, the CFD pictures are already some integral blurring of the original pictures, because the Fourier transform is such integral blurring: each point in the CFD picture contains the information obtained from all points in the original picture.

\section{DISCUSSION}
\label{discussion}

As has been proven in the previous section, using neural nets to determine the types of CFD objects yields excellent results: 
the probability of discrimination in the classification of CFD pictures is close to 100\%.

Apart from the classified object recognition probability (${prob}$), the neural net also gives a series of other useful parameters: the coordinates of the frame bounding the object determined on the picture and the identifier of this frame needed for the separation and subsequent tracking of the object’s details. 
Therefore, if the recognition probability is insufficiently high, then this will imply that the model used describes the picture entering the neural net insufficiently completely and well, i.e., it will be needed to extend and complicate the model and then retrain the neural net. 
On the other hand, if we obtain a sufficiently high probability of the classified object, then it will be possible to extend the model for this object: to partition the available classes into additional subclasses and to retrain the neural net on a new set of classes and subclasses. 
In this way it will be possible to reveal the separate details of the object (class) corresponding to the subclasses of an object with a complex structure on the input picture and so on with new subclasses. 
Nothing of the kind can be done by the standard methods of studying the CFD.

Thus, using neural nets yields much more reliable results than do the standard methods of simulations with several parameters (where the probability of an error is usually higher than that for neural nets by several times). 
However, the main thing is that the use of neural nets is not limited by the number of parameters in the model, i.e., an arbitrarily complex (multiparameter) model of the original picture (and, accordingly, the CFD picture as well) can be used with the same success.

The results of this paper may turn out to be useful to the researchers of the International Event Horizon
Telescope Project:~{\href{https://eventhorizontelescope.org}{https://eventhorizontelescope.org}}.

\section{APPENDIX: Architecture of the YOLOv3 Neural Net}
\label{app1}

The architecture of the YOLOv3 neural net is based on Darknet-53 
(the number in the name means the number of convolutional layers in the architecture).
The architecture uses residual blocks and unsampling and, basically, makes detection in three scales by
dividing the image into $\mbox{13x13, 26x26 and 52x52}$ cells. 
At the output we obtain three tensors 
$\mbox{13x13x(Bx(5+C)), 26x26x(Bx(5+C)) and 52x52x(Bx(5+C))}$, 
where B is the number of frames whose center is in the cell, the frame sizes can be outside the cell, C are the probabilities of classes from the data set (their sum is equal to unity), number 5 denotes five parameters: the probability P that an object was
detected in a given cell, the coordinates of the frame are x and y, its height and width are h and w. 
The number of frames B greater than one is also needed for such situations where the centers of two objects are located
in one cell (for example, a pedestrian against the background of a car). 
For example, for $\mbox{B=2}$ and $\mbox{C=3}$ for each cell in three scales we will obtain 
3549 $\mbox{(13x13+26x26+52x52)}$ vectors 
$\mbox{P1, x1, y1, w1, h1, C1\_1, C1\_2, C1\_3, P2, x2, y2, w2, h2, C2\_1, C2\_2 and C2\_3}$, 
out of which we will filter out only those we need. 
In the YOLOv3 architecture $\mbox{B=3}$. 
Returning to the question about the probability (${prob}$) that the code gives for each object found by it, this is the probability P multiplied by C of the class whose score probability $\mbox{(C1\_1 C1\_2 C1\_3)}$ 
is maximal for a given frame.

\end{document}